\documentclass[a4paper,12pt,english]{article}
\usepackage{amsfonts}
\usepackage{graphicx}
\usepackage{amsmath}
\usepackage{color}

\newcommand{\be}{\begin{equation}}
\newcommand{\ee}{\end{equation}}
\begin{document}

%\tableofcontents

\begin{center}
\noindent{\bf{\LARGE{More on vacua of exotic massive 3D gravity}}}

\smallskip
\smallskip

\smallskip
\smallskip

\smallskip
\smallskip

\smallskip
\smallskip

\noindent{{Gaston Giribet$^1$ and Julio Oliva$^2$}}

\smallskip

\smallskip

\smallskip

\smallskip
\smallskip

\smallskip
\smallskip

\smallskip

\smallskip

\smallskip

\smallskip

{$^1$Departamento de F\'{i}sica, Universidad de Buenos Aires FCEyN-UBA and IFIBA-CONICET} \\ {\it Ciudad Universitaria, pabell\'on 1, 1428, Buenos Aires, Argentina.}
\smallskip

$^2$ {Departamento de F\'{i}sica, Universidad de Concepci\'on} \\ {\it Casilla 160-C, Concepci\'on, Chile.}

\end{center}

\smallskip

\smallskip

\smallskip

\smallskip

\smallskip

\smallskip

\smallskip

\begin{center}
Abstract
\end{center}

In a recent paper \cite{Chernicoff:2018hpb}, we explored the space of solutions of the exotic massive 3D gravity theory proposed in \cite{Ozkan:2018cxj}. We showed that the theory admits a rich space of vacua, including asymptotically Anti de-Sitter (AdS) geometries obeying different types of boundary conditions. The examples include black holes dressed with low decaying gravitons. Based on what happens in other theories of massive gravity, we conjectured that such geometries appear on a curve of the parameter space (chiral curve) where the exotic massive gravity on AdS with sufficiently strong boundary conditions results dual to a 2D chiral conformal field theory. Here, we show that this expectation is consistent with the conserved charges and thermodynamical properties of the black holes of the theory, which have recently been computed \cite{Julio}. When the boundary conditions are relaxed relative to the standard Brown-Henneaux boundary conditions, the theory exhibits solutions consistent with the definition of the so-called Log-gravity. The asymptotic behavior of these solutions presents a logarithmic term in the Fefferman-Graham expansion that, nonetheless, is compatible with the AdS asymptotic symmetries. This long range interaction is due to a mode of the massive gravity that becomes massless precisely on the chiral curve. We construct exact solutions exhibiting this behavior, which admit to be interpreted as fully backreacting gravitational waves propagating on an extremal black hole, and carrying non-vanishing gravitational energy. We also discuss other vacua of the theory, such as Warped-AdS$_3$ black holes, gravitational waves on such backgrounds, AdS$_2\times S^1$ spaces, and black holes in dS$_3$ space.

\smallskip

\smallskip

\smallskip

\smallskip

\smallskip

\smallskip

\smallskip

%\emailAdd{mchernicoff@ciencias.unam.mx}
%\emailAdd{gaston@df.uba.ar}
%\emailAdd{nicolasgrandi@gmail.com} 
%\emailAdd{julioolivazapata@gmail.com}
%\emailAdd{sivasadartantasvueltasnosirve@gmail.com}
%\setlength{\parskip}{8pt}

\newpage

%%%%%%%%%%%%%%%%%

\section{Introduction}

Exotic massive gravity (EMG) is a parity-odd theory describing a propagating massive spin-2 fields in three dimensions (3D) \cite{Ozkan:2018cxj}. It generalizes topologically massive gravity theory (TMG) \cite{TMG} supplementing it with third- and fourth-order terms. EMG can be regarded as the next-to-simplest example of an infinite sequence of consistent 3D models obeying what has been called {\it the third-way mechanism} \cite{thirdway}, i.e. a mechanism by means of which classical metric theories with no action can still satisfy the Bianchi identities on-shell without imposing additional constraints. The simplest example of such a theory is the so-called minimal massive gravity (MMG), proposed in \cite{MMG}; EMG theory is its natural higher-order generalization. 

In the metric formalism, EMG is defined by the field equations
\begin{equation}
R_{\mu \nu }-\frac {1}{2} Rg_{\mu \nu }+\Lambda g_{\mu \nu } + \frac{1}{\mu }C_{\mu \nu}  = T_{\mu \nu }\label{Prima}
\end{equation}
where the Cotton tensor $C_{\mu \nu}$ is 
\begin{equation}
C_{\mu \nu} = \frac 12 \varepsilon_{\mu}^{\ \alpha \beta}\nabla_{\alpha} R_{\beta \nu } +  \frac 12 \varepsilon_{\nu}^{\ \alpha \beta}\nabla_{\alpha} R_{\beta \mu } , \label{Segunda}
\end{equation}
and where the tensor $T_{\mu\nu}$ is
\begin{equation}
T_{\mu\nu}=  \frac{1}{m^2}H_{\mu \nu} - \frac{1}{m^4}L_{\mu \nu } 
\end{equation}
with
\begin{equation}
H_{\mu \nu } = \varepsilon_{\mu }^{\ \alpha \beta}\nabla_{\alpha}C_{\nu \beta } \ , \ \ \ L_{\mu \nu }=\frac{1}{2} \varepsilon_{\mu }^{\ \alpha \beta} \varepsilon_{\nu }^{\ \gamma \sigma} C_{\alpha \gamma} C_{\beta \sigma }.\label{Segundaa}
\end{equation}
$\mu$ and $m$ are two arbitrary coupling constants of mass dimension 1, and $\Lambda $ is the cosmological constant. 

The covariant divergence of $H_{\mu\nu}$ and $L_{\mu\nu}$ does not vanish identically, and this is where the third-way mechanism plays a role: Despite the tensor $T_{\mu\nu}$ not being covariantly conserved identically, one can prove that in virtue of the identities
\begin{equation}
\varepsilon_{\mu}^{\ \alpha \beta }C_{\alpha}^{\sigma}(R_{\beta\sigma}-\frac 12 Rg_{\beta\sigma}-\frac{1}{m^2} H_{\beta\sigma} )=\varepsilon_{\mu}^{\ \alpha \beta }C_{\alpha }^{\sigma}(-\Lambda g_{\beta\sigma} -\frac{1}{m^4} L_{\beta\sigma})=0\label{This}
\end{equation}
the conservation condition $\nabla_{\nu}T^{\nu}_{\ \mu}=0$ does hold on-shell without imposing incompatible constraints. This is the type of theories recently studied in \cite{thirdway, thirdway2}, for which the {\it matter} contribution $T_{\mu\nu}$ that supplements TMG field equations satisfies the Bianchi identities as a consequence of the gravitational field equation itself.

In a recent paper \cite{Chernicoff:2018hpb}, we explored the space of solutions to the field equations (\ref{Prima})-(\ref{Segundaa}) and we showed that it contains interesting geometries, including asymptotically AdS$_3$ black holes and gravitational waves. We also showed that, on a particular curve of the parameter space, the theory also admits logarithmically decaying solutions in AdS$_3$. These are analogous to the solutions of TMG that appear at the so-called chiral point \cite{Chiral}. Therefore, we conjectured that, on a special curve of the parameter space (see (\ref{chiral}) below), EMG on AdS$_3$ with sufficiently strong boundary conditions is dual to a 2D chiral conformal field theory (CFT$_2$). Here, we will show that this conjecture is consistent with the conserved charges and thermodynamical properties of the black holes of the theory \cite{Julio}. 
In section 2, we will study the simplest black hole solution of the theory, the Ba\~{n}ados-Teitelboim-Zanelli (BTZ) solution, which describe black holes in AdS$_3$ space. We will review the main properties of BTZ, now seen as a solution to EMG, and in particular its conserved charges. By studying the black hole thermodynamical properties, and assuming the existence of a CFT$_2$ dual description, in section 3 we will obtain the central charges of the boundary theory. Then, in section 4, we will study the special properties the theory exhibits when formulated on the aforementioned special curve of the parameter space, to which hereafter we will refer to as the {\it chiral curve}. On this curve, the left-moving central charge of the dual CFT$_2$ vanishes and this is taken as evidence that, provided sufficiently strong AdS$_3$ boundary conditions are prescribed, the dual CFT$_2$ turns out to be chiral. In contrast, when weakened AdS$_3$ boundary conditions are considered, the dual CFT$_2$ has both left- and right-moving dynamics, and this is in correspondence with additional normalizable massless fields appearing in the bulk. In section 5, we will discuss explicit examples of these geometries, which describe gravitational waves propagating on extremal black holes. Sections 6-10 will be dedicated to study other vacua of the theory, including massive gravitational waves in AdS$_3$ and other deformations of the extremal BTZ solution, black holes in Warped-AdS$_3$ (WAdS$_3$) space, evanescent gravitational waves on WAdS$_3$ space, AdS$_2\times S^1$ spaces, and hairy black holes in both AdS$_3$ and dS$_3$ space. 

\section{BTZ black holes}

As other {\it bona fide} massive modifications to Einstein theory in 3D, EMG admits the BTZ black hole as exact solution. Its metric takes the form \cite{BTZ}
\begin{equation}
ds^2_{M,J} = -N^2(r)dt^2 +\frac{dr^2}{N^2(r)}+ r^2 (N^{\varphi }(r)dt+ d\varphi )^2\label{BTZ}
\end{equation}
with
\begin{equation}
N^2(r) = \frac{r^2}{\ell^2}-8GM+\frac{16G^2J^2}{r^2}\ , \ \ \ N^{\varphi }(r)=-\frac{4GJ}{r^2}\label{BTZ2}
\end{equation}
where $t\in \mathbb{R}$, $r\in \mathbb{R}_{\geq 0}$, $0\leq \varphi \leq 2\pi $, and where $\ell^2=-1/\Lambda$. $M$ and $J$ are two arbitrary integration constants. For generic $M$, the space (\ref{BTZ})-(\ref{BTZ2}) has isometry $\mathbb{R}\times SO(2)$. This geometry represents a black hole in AdS$_3$ space, provided $M\geq 0$ and $|J|\leq M\ell $. The horizon radii are
\begin{equation}
r_{\pm } = \sqrt{2G\ell(\ell M+J)}\pm \sqrt{2G\ell(\ell M-J)},
\end{equation}
with $0\leq r_- \leq r_+$. This means
\begin{equation}
M=\frac{r_+^2+r_-^2}{8G\ell^2} \ , \ \ \ J=\frac{r_+r_-}{4G\ell}.
\end{equation}

Being a 3D Einstein spaces with negative cosmological constant, BTZ black holes are locally equivalent to AdS$_3$, and they can be constructed by making global identifications on the universal covering of the maximally symmetric geometry \cite{BTZ2}. It is convenient to define the geometric temperatures 
\begin{equation}
T_{\pm}=\frac{r_+\pm r_-}{2\pi \ell^2}\label{La10}
\end{equation}
which are the inverse of periods of the geometry \cite{BTZ2}. In terms of these, the Hawking temperature is given by
\begin{equation}
T_{\text{H}}=2\ \frac{ T_+T_-}{T_+ +T_-}=\frac{r^2_+ - r^2_-}{2\pi \ell^2 r_+}.
\end{equation}

The conserved charges of the BTZ solution in EMG were recently computed in \cite{Julio} (see also \cite{Tekin}), being the energy
\begin{equation}
Q[\partial_t] = \frac{1}{8G\ell^2}\Big[ \Big( 1-\frac{1}{m^2\ell^2}\Big) (r^2_++r^2_-)-\frac{2}{\mu\ell }r_+r_-\Big]\label{Q1}
\end{equation}
and the angular momentum
\begin{equation}
Q[\partial_{\varphi }] = \frac{1}{8G\ell }\Big[ \Big( 1-\frac{1}{m^2\ell^2}\Big) 2r_+r_- -\frac{1}{\mu \ell }(r^2_++r^2_-) \Big].\label{Q2}
\end{equation}
These are the conserved charges associated to the Killing vectors $\partial_t$ and $\partial_{\varphi }$, respectively. With them, one obtains the black hole entropy 
\begin{equation}
S_{\text{BH}}=\frac{\pi }{2G}\Big[ \Big( 1-\frac{1}{m^2\ell^2}\Big)  r_+ -\frac{1}{\mu \ell } r_-\Big]\label{TheS}
\end{equation}
which follows from the first principle of black hole thermodynamics
\begin{equation}
dQ[\partial_{t }] = T_{\text{H}}  dS_{\text{BH}} - \Omega \ dQ[\partial_{\varphi}]
\end{equation}
with the horizon angular velocity $\Omega = N^{\varphi }(r_+)=-{r_-}/{r_+}$. Notice that, while for $m^2=\mu =\infty $ (\ref{TheS}) obeys the Bekenstein-Hawking area law, the presence of higher-derivative terms introduce corrections to it. This is, of course, a common feature of higher-curvature gravity. 

\section{Central charge}

As for other 3D models, we can rewrite the entropy (\ref{TheS}) in a suggestive form: A Cardy-type formula of the dual CFT$_2$ can be written in terms of the geometric temperatures (\ref{La10}) and the entropy (\ref{TheS}), namely \cite{Strominger}
\begin{equation}
S_{\text{BH}}=\frac{\pi^2\ell }{3} \Big( c_+ T_+ + c_- T_-\Big)\label{Cardy2}
\end{equation}
which is obeyed for the following values of the central charge(s)
\begin{equation}
c_{\pm } = \frac{3\ell }{2G} \Big( 1-\frac{1}{m^2\ell^2}\mp\frac{1}{\mu\ell}\Big).\label{c}
\end{equation}
This, of course, agrees with the right-moving and left-moving central charges of TMG in the limit $m^2\to \infty$.  

Now, let us convince ourselves of (\ref{c}) being the correct result: We can write the operators $L_0$ and $\bar{L}_0$, which generate the $SO(2)\times \mathbb{R}$ piece of the global conformal group at the cylindrical boundary ($r=\infty $). These operators would correspond to the zero-modes of the Virasoro operators $L_n$, $\bar{L}_n$ that generate the whole asymptotic symmetry algebra \cite{BH}. These are related to the conserved charges (\ref{Q1})-(\ref{Q2}) as follows
\begin{equation}
L_0+\bar{L}_0 = \ell Q[\partial_{t}] \ , \ \ \ L_0-\bar{L}_0 =  Q[\partial_{\varphi }]  .
\end{equation}
Denoting $L^+_0\equiv L_0$ and $L^-_0\equiv \bar{L}_0$, we can write
\begin{equation}
L^{\pm}_0 = \frac{c_{\pm}}{24}\frac{(r_+\pm r_-)^2}{\ell^2}.\label{losL}
\end{equation}
Notice that these values are consistent with (\ref{c}) and with the vacuum (vac) of the theory: In fact, global AdS$_3$ space corresponds to (\ref{BTZ})-(\ref{BTZ2}) with $r^2_+=-\ell^2$ and $r_-=0$. Then, the values of the charges (\ref{Q1})-(\ref{Q2}) corresponding to the vacuum geometry are
\begin{equation}
Q^{\text{(vac)}}[\partial_{t}]=\frac{1}{8G}\Big( \frac{1}{m^2\ell^2}-1\Big) \ , \ \ \  Q^{\text{(vac)}}[\partial_{\varphi }]=\frac{1}{8G\mu},
\end{equation}
and this implies that, for the vacuum,
\begin{equation}
L^{\pm \text{(vac)}}_0 = -\frac{c_{\pm}}{24}.
\end{equation}
Moreover, notice that with (\ref{losL}) we can alternatively write the Cardy formula as follows
\begin{equation}
S_{\text{CFT}}=4\pi \sqrt{L^+_0\ \frac{ c_+}{24}} + 4\pi \sqrt{L^-_0\ \frac{ c_-}{24}} \label{Cardy1}
\end{equation}
which, again, reproduces the black hole entropy (\ref{TheS}). In other words, we get $S_{\text{BH}}=S_{\text{CFT}}$ and thus (\ref{c}) seem to be the correct values of the central charges. 

Conformal and diffeomorphism anomalies in the dual CFT$_2$ pose restrictions on the parameters of the bulk theory for the latter to have chances of being well defined \cite{Witten}: On the one hand, assuming the existence of a dual CFT$_2$ with an $SL(2,\mathbb{C})$ invariant vacuum, the values of the central charges $c_{\pm }$ imply that the theory cannot be defined for continuous values of $3\ell(m^2\ell^2 -1)/(2Gm^2\ell^2)$. This follows from the Zamolodchikov $c$-theorem in the dual CFT$_2$. On the other hand, as in TMG, the diffeomorphism anomaly in the dual CFT$_2$ is controlled by the number $c_+-c_-=-3/(G\mu )$, which has to take values $1/(8G\mu) \in \mathbb{Z}$ for the theory to be modular invariant. 

\section{Chiral curve}

In \cite{Chernicoff:2018hpb}, we were particularly interested in the curve of the parameter space of EMG defined by the relation
\begin{equation}
\mu =\frac{m^2\ell}{1-m^2\ell^2} \ , \ \ \ \Lambda=-\frac{1}{\ell^2} \label{chiral}\ .
\end{equation}
Hereafter, we will refer to this as the chiral curve, as the conjecture in \cite{Chernicoff:2018hpb} was that EMG with couplings obeying (\ref{chiral}) and sufficiently strong boundary conditions turns out to be dual to a chiral CFT$_2$. Indeed, when the values of the central charge (\ref{c}) are evaluated on (\ref{chiral}), we obtain
\begin{equation}
c_-=0 \ , \ \ \ c_+=\frac{3\ell }{G}\Big( 1-\frac{1}{m^2\ell^2}\Big),\label{este}
\end{equation}
which is a necessary condition for chirality \cite{Chiral}. This is related to the fact that logarithmically decaying modes in EMG appear at (\ref{chiral}), and chirality demands these modes to be decoupled. The picture might be similar to what happens in TMG. I.e. at the point $c_-=0$, two theories exist: while the theory with standard boundary conditions in AdS$_3$  \cite{BH} results dual to a chiral CFT$_2$, the theory defined with weaker boundary conditions is presumably dual to a logarithmic CFT$_2$ \cite{LogGravity}. 

%EMG theory not admitting a variational principle in the metric formulation\footnote{The theory does admit a Chern-Simons like action when written in terms of the dreibein and the spin connection.}, to obtain the central charge of the dual CFT one cannot resort to the standard procedure involving the action. It demands to resort to the conserved charge computation \cite{Julio}..... The central charge (\ref{c}) has a form similar to the closely related MMG theory, which is defined by replacing in the field equations (\ref{Prima}) the tensor $T_{\mu \nu }$ by the second-order contribution
%\begin{equation}
%T_{\mu \nu } = 2R_{\mu}^{\ \rho}R_{\rho \nu}-\frac{3}{2}RR_{\mu\nu}-g_{\mu\nu}\Big( R_{\rho \sigma}R^{\rho \sigma}-\frac{5}{8} R^2\Big)}.
%\end{equation}

The theory exhibits other special features at $c_-=0$. At this point, all BTZ solutions satisfy the extremal condition 
\begin{equation}
Q[\partial_{\varphi}] = \ell Q[\partial_{t}], 
\end{equation}
regardless the values of $M$ and $J$. This suggests a high degeneracy of the spectrum.  Also at this point, the solutions with $r_-=r_+$ (i.e. $|J|=\ell M$) present vanishing entropy, $S_{\text{BH}}=0$. We will see in the next section that, in addition to BTZ black holes, at $c_-=0$ the theory presents a large family of asymptotically AdS$_3$ solutions with $\mathbb{R}\times SO(3)$ isometries, vanishing entropy, and non-vanishing gravitational energy.

\section{Chiral waves on black holes}\label{La5}

The metric of the extremal BTZ solution takes the form
\begin{equation}
ds^{2}_{M,\mp \ell M }=-\frac{\left(  r^{2}-r_{+}^{2}\right)  ^{2}}{\ell^2 r^{2}}dt^{2}+\frac
{\ell^2r^{2}}{\left(  r^{2}-r_{+}^{2}\right)  ^{2}} dr^{2} +\frac{r^{2}}{\ell^2}\left( \ell d\varphi \pm \frac
{r_{+}^{2}}{r^{2}}dt\right)^{2}\ ,\label{eBTZ}
\end{equation}
where $r_+^2=r_-^2=4\ell^2 {GM}$. Starting from this geometry, a new family of non-locally AdS$_3$ solutions to EMG can be constructed by means of a Kerr-Schild transformation: At the chiral point (\ref{chiral}), the following deformation of the extremal BTZ is also a solution
\begin{equation}
d{s}^{2}\equiv {g}_{\mu \nu }dx^{\mu }dx^{\nu }=ds^{2}_{M,-\ell M}+L\left(  t,r,\varphi\right)
\left(  dt+\ell d\varphi\right)  ^{2}\ ,\label{solutiona}
\end{equation}
with
\begin{equation}
L\left(  t,r,\varphi\right) = A(x^+)\log (r^2-r_+^2)+B(x^+)(r^2-r_+^2)^{\frac{1+m^2\ell^2}{2}}+C(x^+)(r^2-r_+^2)+D(x^+)\label{modes}
\end{equation}
where $x^{\pm}=t\pm \ell\varphi $, and where $A$, $B$, $C$, and $D$ are arbitrary functions that depend only on $x^+$; namely, they satisfy
\begin{equation}
\partial_- A = 0 \ , \ \ \ \partial_- B = 0 \ , \ \ \ \partial_- C = 0 \ , \ \ \ \partial_- D = 0 \ , \ \ \ 
\end{equation}
with $\partial _{\pm}=\partial / \partial x^{\pm}$.

We can now analyze all the contributions appearing in (\ref{modes}) and study their physical meaning: Let us start with the $A$-modes, which accompany the logarithm in (\ref{modes}). This logarithmically growing potential is compatible with the AdS$_3$ boundary conditions
\begin{eqnarray}
{g}_{tt}&\simeq & \frac{r^2}{\ell^2} + \mathcal{O}(\log r ) \ , \ \ \ {g}_{t\varphi }\simeq  \mathcal{O}(\log r ) \ , \ \ \ {g}_{\varphi \varphi }\simeq  {r^2} + \mathcal{O}(\log r ) \ , \nonumber \\
{g}_{rr}&\simeq & \frac{\ell^2}{r^2} + \mathcal{O}( 1/r^4 ) \ , \ \ \ {g}_{r\varphi }\simeq  \mathcal{O}(1 ) \ , \ \ \ {g}_{rt }\simeq  \mathcal{O}(1 ) \ , \label{asymptotig}
\end{eqnarray}
which are the boundary conditions proposed in \cite{GJ} in the context of Log-Gravity \cite{LogGravity}. Periodicity of $\varphi $ implies functions $A$ to be periodic under $t\to t+2\pi \ell $. This yields the mode expansion
\begin{equation}
A(x^+) = \sum_{n=0}^{\infty } \Big( \zeta_n \ e^{in({t/\ell +\varphi})} + \zeta_n^* \ e^{-in({t/\ell +\varphi })}\Big).
\end{equation} 
The periodicity in $t$ is well understood, since $\ell $ is the time that takes for a massive or massless mode in AdS$_3$ to go towards the boundary and bounce back due to the gravitational potential. The long range interaction represented by the $A$-mode is due to the fact that this mode is actually a mode of the massive theory that becomes massless precisely on the curve (\ref{chiral}). In fact, it satisfies the fourth-order {\it massless} wave equation
\begin{equation}
\Box^2_{A} \ e^{in(t/\ell + \varphi )} \log (r^2-r^2_+)= 0,\label{box}
\end{equation} 
where $n\in\mathbb{Z}$ and where $\Box_A $ is the Laplacian operator corresponding to the metric (\ref{solutiona})-(\ref{modes}) with a given function $A$. The fact that the $A$-mode is massless, in the sense of obeying (\ref{box}), does not mean that the full non-linear solution has vanishing gravitational energy. In fact, we will see below that the $A$-mode does contribute to the gravitational energy of the geometry. 

The $B$-mode also satisfies a wave equation: it obeys the massive second order equation
\begin{equation}
(\Box_A - \kappa ^2) e^{in(t/\ell + \varphi )} (r^2-r^2_+)^{\frac{1+m^2\ell^2}{2}}=0 \ , \ \ \ \text{with} \ \ \kappa^2\ell^2={(m^2\ell^2+3)(m^2\ell^2+1)}.\label{boxqe}
\end{equation}
However, unlike the $A$-mode, the $B$-mode does not represent a massive mode in AdS$_3$ as it grows like $\sim r^{1+m^2\ell^2}$ and thus violates the asymptotic (\ref{asymptotig}). In other words, its linearized version would correspond to a non-normalizable state in AdS$_3$. To understand the origin of the $B$-mode, we can notice that in the limit $m^2\to 0$ equation (\ref{chiral}) implies $\mu =0$, and this is the point of the parameter space where the field equations are satisfied by any conformally flat metric. This is because at $m^2=\mu =0$ all the terms in the field equations contain the Cotton tensor and its derivatives. In that limit, the $B$-mode behaves at large distance like $\sim r$, and this is the typical behavior of the mode of conformal gravity \cite{julioconformal}. We will discuss non-linear solutions associated to this mode in section \ref{La10}.

While the $A$- and $B$-modes are lawful modes of the massive theory, the $C$- and the $D$-modes correspond to general relativity (GR) modes and they can be regarded as pure gauge. Therefore, we can set $C=D=0$. When $m^2\ell^2=1$, the $B$- and the $C$-modes degenerate. The solutions with $B=C=0$ and generic $A$ and $D$ represent asymptotically AdS$_3$ solutions, which nonetheless are not locally equivalent to AdS$_3$ and are not conformally flat. In the context of AdS/CFT correspondence, function $A$ acts as a source of a field in the dual CFT$_2$, while it determines the boundary condition for the bulk field configuration. 

%Primero, excluimos los modos: El $A$-mode ... The $B$-mode ... when $m^2=0$, the $B$-mode corresponds to the conformal mode, which goes at large distance linearly, $\sim r$. In the case $m^2\ell^2=1$ the $B$-modes and the $C$-modes degenerate.   ... This is because, provided (\ref{chiral}), $m^2=0$ corresponds to $\mu=0$, which is the limit in which the theory becomes conformal gravity. Since $m^2\ell^2\geq 0$, the $B$ violate the asymptotically AdS$_3$ boundary conditions, so we fixe $B=0$. 

The particular case $M=J=0$ of solution (\ref{solutiona})-(\ref{modes}) corresponds to the so-called AdS-waves, which are special example of Siklos spacetimes \cite{TMGwaves, NMGwaves}; see also \cite{asteriscoA, asteriscoZ}. Metrics (\ref{solutiona})-(\ref{modes}) thus generalize the TMG wave solutions of \cite{TMGwaves}, and then can be interpreted as non-linear gravitational waves on the extremal BTZ geometry. Function $A$ determines the profile of the wave. Remarkably, equation (\ref{box}) is also obeyed if one replaces the Laplacian operator $\Box_A$ by the Laplacian of the undeformed BTZ solution, $\Box_{A=0}$. --This is the case even when the two operators do not identically coincide.-- Therefore, (\ref{solutiona})-(\ref{modes}) can either be interpreted as gravitational waves propagating in stealth mode on the black hole background, or as non-linear waves propagating on their own backreacting geometries.

As explained above, the periodicity in $t$ is easily explained because $\ell $ is the characteristic time scale in AdS space. That is, the period of the wave is determined by the AdS radius itself and not by the rotation period of the black hole. What the black hole angular momentum does determine is the growth of the radial dependence close to the horizon. It produces a divergence in the angular and temporal components of the metric at $r=r_+$, and this ultimately determines the direction of the rotation: The wave turns out to be co-rotating with the black hole.  

Solution (\ref{solutiona})-(\ref{modes}) also comes to generalize the Log-gravity solution presented in \cite{Garbarz:2008qn}, which corresponds to the particular case $A=const$. The gravitational energy and angular momentum of (\ref{solutiona})-(\ref{modes}) can be computed, yielding
\begin{equation}
Q[\partial_t] = \frac{1}{4\pi G \ell} \Big( 1+\frac{1}{m^2\ell^2}\Big)\int_0^{2\pi \ell} A(s) ds\ , \ \ \ \ Q[\partial_{\varphi }] = -\ell Q[\partial_t],
\end{equation}
which reduces to the result of \cite{Julio} in the particular case $\partial_{\pm }A=0$. 

\section{Massive waves in the mirror}

A mirror image of the family of solutions (\ref{solutiona})-(\ref{modes}) can be constructed by means of a similar Kerr-Schild ansatz but considering as a seed the extremal BTZ with the opposite angular momentum. This anti-clockwise version of (\ref{solutiona})-(\ref{modes}), however, does not behave in the same way as the theory is not parity invariant. In fact, the following deformation of the $M=+J/\ell $ black hole is also a solution of EMG when (\ref{chiral}) holds,
\begin{equation}
d{s}^{2}=ds^{2}_{M,+\ell M}+L\left(  t,r,\varphi\right)
\left(  dt-\ell d\varphi\right)  ^{2}\ ,\label{sola}
\end{equation}
with
\begin{equation}
L\left(  t,r,\varphi\right) = A(x^-)(r^2-r_+^2)\log (r^2-r_+^2)+B(x^-)(r^2-r_+^2)^{\frac{1-m^2\ell^2}{2}}+C(x^-)(r^2-r_+^2)+D(x^-),
\label{TY}
\end{equation}
and where the functions now satisfy
\begin{equation}
\partial_+ A = 0 \ , \ \ \ \partial_+ B = 0 \ , \ \ \ \partial_+ C = 0 \ , \ \ \ \partial_+ D = 0 \ . \ \ \ 
\end{equation}

Again, at $m^2=0$ the $B$-mode corresponds to the conformal mode, and at $m^2\ell^2=1$ the $B$-mode degenerates with the $C$-mode. Within the range $0\leq m^2\ell^2 <1$ the $B$-mode blows up at infinity. In contrast, when $m^2\ell^2 >1$, the $B$-modes becomes compatible with the Brown-Henneaux AdS$_3$ boundary conditions \cite{BH}. Indeed, provided $A=C=0$, the solution (\ref{sola}) obeys
\begin{eqnarray}
{g}_{tt}&\simeq & \frac{r^2}{\ell^2} + \mathcal{O}(1 ) \ , \ \ \ {g}_{t\varphi }\simeq  \mathcal{O}(1 ) \ , \ \ \ {g}_{\varphi \varphi }\simeq  {r^2} + \mathcal{O}(1 ) \ , \nonumber \\
{g}_{rr}&\simeq & \frac{\ell^2}{r^2} + \mathcal{O}( 1/r^4 ) \ , \ \ \ {g}_{r\varphi }\simeq  \mathcal{O}(1/r^3 ) \ , \ \ \ {g}_{rt }\simeq  \mathcal{O}(1/r^3 ) \ , \label{BHaa}
\end{eqnarray}
and the $B$-mode now does represent a massive mode in AdS$_3$. It satisfies the wave equation
\begin{equation}
(\Box_A - \kappa ^2) e^{in(t/\ell - \varphi )} (r^2-r^2_+)^{\frac{1-m^2\ell^2}{2}}=0 \ , \ \ \ \text{with} \ \ \kappa^2\ell^2={(m^2\ell^2-3)(m^2\ell^2-1)};\label{boxq}
\end{equation} 
it decays rapidly enough at large distance and acts as a source in the dual CFT$_2$.

\section{Warped black holes}\label{La7}

Now, let us discuss black holes in warped deformations of AdS$_3$ space (WAdS$_3$). These solutions were found in \cite{Clement} in the context of TMG, and they were exhaustively studied in \cite{Anninos:2008fx}. The metric of these black holes is given by
\begin{eqnarray}
ds^2_{r_+ , r_-}&=&dt^2 + \frac{\ell^2}{\nu^2+3} \frac{dr^2}{(r-r_+)(r-r_-)}-2\Big( \nu r  +\frac 12 \sqrt{r_+r_-(\nu^2+3)}\Big) dtd\phi \nonumber \\
&&+\frac{r}{4}\Big( 3(\nu^2-1)r+(\nu^2+3)(r_++r_-)+4\nu \sqrt{r_+r_-(\nu^2+3)}\Big)d\phi^2\label{WAR}
\end{eqnarray}
where $t\in \mathbb{R}$, $r\in \mathbb{R}_{\geq 0}$, and $0\leq \phi \leq 2\pi $; $\ell $ is a real parameter related to the Gaussian curvature of the manifold; and $r_{\pm}$ are two integration constants taken to satisfy $0\geq r_- \geq r_+$. $r_+$ represents the outer horizon of the black hole, and $r_-$ represents its inner horizon.

WAdS$_3$ black holes (\ref{WAR}) are locally equivalent to the three-dimensional section of the G\"odel cosmological solution to Einstein equations. This is locally equivalent to either squashed ($\nu^2<1$) or stretched ($\nu^2>1$) deformations of AdS$_3$, with the parameter $\nu $ controlling such deformation. $\nu^2 =1$ corresponds to locally AdS$_3$ (undeformed) spaces. $\nu =0 $ corresponds to another interesting case that we will discuss below: it is locally AdS$_2\times S^1$. In fact, space (\ref{WAR}) and its equivalents can be written as a Hopf fibration over AdS$_2$ or over its Euclidean version $\mathbb{H}_2^+$, with the fiber being $\mathbb{R}$. $\nu =0$ is a kind of extreme limit.

Black holes (\ref{WAR}) are solutions to EMG field equations provided
\begin{equation}
\mu=\frac{\ell^3m^4\nu}{3(\nu^4-\nu^2(1+\ell^2m^2)+\ell^4m^4/9)} \ , \ \ \nu^6-2\nu^4+\nu^2+m^4\ell^2 (1+\Lambda \ell^2)/9=0\label{atorranta}.
\end{equation}
They have non-vanishing gravitational energy, which is given by
\begin{equation}
Q[\partial_t] = \frac{(\nu^2+3)(3\nu^2+\ell^2m^2)}{24G m^2\ell^3}\Big( r_+ + r_- -\frac{1}{\nu} \sqrt{(\nu^2+3)r_-r_+}\Big).
\end{equation}
Notice that this formula generalizes the TMG formula to which it reduces in the limit $m^2\to \infty$.

\section{Evanescent gravitons in WAdS$_3$}\label{La8}

EMG field equations admits further deformations of (\ref{WAR}). In particular, it admits solutions that describe evanescent gravitational waves on WAdS$_3$ spaces. To see this, let us consider the case $r_{\pm }=0$, whose metric is
\begin{equation}
ds^{2}_{0,0}=dt^{2}+\frac{2dr^{2}}{\left(  \nu^{2}+3\right)  r^{2}}-2\nu
rdtd\phi+\frac{3}{4}r^{2}\left(  \nu^{2}-1\right)  d\phi^{2},
\end{equation}
where we are using the convention $\ell \equiv 1$ for short. Field equations (\ref{Prima})-(\ref{Segundaa}), provided equations (\ref{atorranta}) hold, admits the following solution
\begin{equation}
ds^2=ds^{2}_{0,0}+e^{-\omega
t}r^{-\frac{2\nu\omega}{\nu^{2}+3}}A(x^-) (dx^-)^2  \ ,\label{defo}
\end{equation}
where $A$ is an arbitrary periodic function of $x^-$, with
\begin{equation}
x^{\pm}=\phi\pm \frac{2}{\left(  \nu^{2}+3\right)  r}\ ,
\end{equation}
and where $\omega$ is a parameter given by the polynomial equation 
% \left(  d\phi+\frac{2dr}{\left(  \nu^{2}+3\right)  r^{2}}\right)  ^{2}
\begin{equation}
(5\nu^{2}+6\nu\omega+\omega^{2}+3)m^{4}-3\nu(\nu+\omega)(5\nu^{2}+6\nu
\omega+\omega^{2}+3)m^{2}-27\nu^{2}(\nu^{2}-1)(\nu^{2}-3\nu\omega-\omega
^{2}-3)=0 .\label{Poli}
\end{equation}
This solution generalizes one of the TMG solutions found in \cite{HenneauxTroncoso}. In fact, notice that in the limit $m^2\to \infty$, (\ref{Poli}) reduces to the TMG condition $\omega =-3\pm \sqrt{4\nu^2-3}$ found therein. Provided $\omega >0$, solution (\ref{defo}) describes an evanescent gravitational wave on the massless WAdS$_3$ black hole geometry. As the Log-gravity waves discussed in section \ref{La5}, waves (\ref{defo}) concentrate close to the horizon (in this case, at $r=0$), where the metric components $g_{tt}$, $g_{\varphi t}$, and $g_{\varphi \varphi}$ diverge (assuming $\omega \nu >0$, which is necessary for the solution to be asymptotically WAdS$_3$). This wave satisfies the massive equation
\begin{equation}
\Big(\Box_A -\kappa^2 \Big) e^{-\omega t +inx^- }r^{-\frac{2\nu\omega}{\nu^{2}+3}} =0 \ , \ \ \ \text{with} \ \ \kappa^2 = \omega (\omega - 2\nu ),
\end{equation}
$n\in \mathbb{Z}$. $\Box_A$ stands here for the Laplacian corresponding to metric (\ref{defo}); nevertheless, the same equation holds if one replaces $\Box_A$ by $\Box_{A=0}$. This follows from the arbitrariness of $A$, what makes the {\it ansatz} to be linear in the function of $x^-$. Perturbation (\ref{defo}) carries no gravitational energy.

\section{AdS$_2 \times S^1$ vacua}

EMG generalizes TMG and, consequently, contains conformal gravity as a particular case: The latter corresponds to the limit $\mu \to 0 $, $m^2\to \infty$. In particular, AdS$_2 \times S^1$ and $\mathbb{H}_2^+ \times \mathbb{R}$ spaces appear as solutions of EMG only in that limit. Let us briefly discuss these geometries and relate them with the ones discussed above: In fact, an ingenious trick to work out the AdS$_2\times S^1$ case is to think of it as the $\nu \to 0$ limit of the WAdS$_3$ space. From (\ref{atorranta}), one observes that when $\nu$ goes to zero $\mu$ also goes to zero, in such a way that
\begin{equation}
\frac{\mu}{\nu}=\frac{3}{\ell} \ , \ \ \ \ \ell^2=-\frac{1}{\Lambda}. \label{cerapio}
\end{equation}

We can think of the large $m^2 $ limit. In that case, the left- and right-moving central charges for asymptotically WAdS$_3$ spaces are \cite{Compere}
\begin{equation}
c_- = \frac{4\nu \ell}{G(\nu^2+3)} \ , \ \ \ \ c_+ = \frac{(5\nu^2+3)\ell }{G\nu (\nu^2+3)} .\label{oty}
\end{equation}
Notice that these expressions are consistent with (\ref{c}) in the AdS$_3$ limit $\nu \to 1$: Considering $\mu \ell=3$ and $m^2=\infty $, one obtains $c_-=\ell/G$ and $c_+=2\ell/G$, in accordance with (\ref{c}). In the AdS$_2$ limit $\nu \to 0$, on the other hand, one gets
\begin{equation}
c_- = 0 \ , \ \ \ \ c_+ = \frac{3}{G\mu } \equiv 24k,\label{otyy}
\end{equation}
where the Chern-Simons (CS) level $k$ relates to the TMG coupling constant by $k/(4\pi)= 1/(32\pi G \mu)$, which is the correct normalization of the gravitational CS action 
\begin{equation}
I_{\text{CS}} = \frac{k}{4\pi } \int d^3x \ \varepsilon^{\mu \nu \rho} \Big( \Gamma_{\mu \alpha}^{\beta} \partial_{\nu} \Gamma_{\rho \beta }^{\alpha} +\frac{2}{3} \Gamma_{\mu \alpha}^{\eta} \Gamma_{\nu \beta}^{\alpha} \Gamma_{\rho \eta}^{\beta}\Big)
\end{equation}
that yields the field equation $(1/\mu)C_{\mu\nu}=0$, in agreement with the $\mu\to 0 $ limit of (\ref{Prima}). It is worth noticing that (\ref{otyy}) exactly reproduces the result of \cite{AdS2} for $\mathbb{H}_2^+$ holography. Furthermore, one can easily verify that the non-perturbative states discussed in \cite{AdS2} are actually diffeomorphic to the $\nu \to 0$ limit of the metric (\ref{WAR}), which have vanishing conserved charges.

\section{Hairy black holes in (A)dS}\label{La10}

The theory (\ref{Prima}) can also be defined in the limit $m^2\to 0$, $\mu \to 0$, in which the Einstein tensor disappears from the field equations. This limit, however, is subtle; it has to be taken in the right order for the consistency equations $\nabla_{\mu}T^{\mu}_{\ \nu}=0$ to hold. When Einstein tensor is absent, the field equations are satisfied for all conformally flat spaces. This is because the Cotton tensor is zero if and only if a metric is conformally equivalent to flat space, at least locally. A very interesting case of conformally flat space is the following \cite{julioconformal}
\begin{equation}
ds^2= -\frac{(r-r_+)(r-r_-)}{\ell^2} dt^2 + \frac{\ell^2 dr^2}{(r-r_+)(r-r_-)}+r^2d\varphi^2\label{hairy}
\end{equation}
with $t\in \mathbb{R}$, $r\in \mathbb{R}_{\geq 0}$, and $0\leq \varphi \leq 2\pi $. This space, provided either $r_+$ or $r_-$ is positive, represents a static black hole. For $\ell^2>0$, (\ref{hairy}) is asymptotically AdS$_3$, in the sense of the weakened boundary conditions studied in \cite{Julio2}. In particular, this implies the behavior
\begin{equation}
g_{tt}\simeq -\frac{r^2}{\ell^2}+\mathcal{O}(r).
\end{equation}
This linear term in $r$, together with its analogs in the angular components of the metric, correspond to the conformal gravity mode that, as we discussed in section \ref{La5}, appears in the $m^2\to 0 $ limit. Despite this weakened asymptotics, solution (\ref{hairy}) has finite conserved charges: Defining
\begin{equation}
M\equiv -\frac{r_+ r_-}{8G\ell^2 } \ , \ \ \ \ N=\frac{(r_+ + r_-)^2}{32{G}\ell^2}
\end{equation}
we obtain that the mass and the angular momentum are given by
\begin{equation}
Q[\partial_t]= -\frac{M+N}{m^2\ell^2}=-\frac{(r_+-r_-)^2}{32Gm^2\ell^4}    \ , \ \ \ \ Q[\partial_{\varphi}]= -\frac{M+N}{\mu}= -\frac{(r_+-r_-)^2}{32G\mu\ell^2},
\end{equation}
respectively. These values for the charges can be obtained from the results (\ref{Q1}) and (\ref{Q2}) in a simple way: One considers in (\ref{hairy}) the coordinate changes $r\to r-(r_++r_-)/2$, and then take in (\ref{Q1})-(\ref{Q2}) the limit $m^2\to 0$, $\mu \to 0$.

Assuming $r_+\geq r_-\geq 0$, for which two horizons exist, the temperature of the black hole takes the form
\begin{equation}
T_{\text{H}} = \frac{r_+-r_-}{4\pi \ell^2},\label{ttt}
\end{equation}
which vanishes in the extremal limit $r_+=r_-$, for which the near horizon symmetry renders AdS$_2\times S^1$. This means that the entropy of the black hole is proportional to the difference between the outer and inner horizons. More precisely, 
\begin{equation}
S=-\frac{\pi (r_+-r_-)}{4Gm^2\ell^2 },\label{sss}
\end{equation}
which follows from the first principle of the black hole thermodynamics. The negative sign in (\ref{sss}) and in the conserved charges demands to perform $m\to im$. Remarkably, this entropy can be obtained from (\ref{ttt}) by using Cardy formula: By considering the large $1/m^2$, $1/\mu $ limit in (\ref{c}), we actually obtain
\begin{equation}
S=\frac{\pi^2\ell }{3 }(c_+ + c_-)T_{\text{H}}.
\end{equation}

This black hole solution generalizes the static BTZ black hole solution, which corresponds to the special case $r_+=-r_-\geq 0$ (i.e. $N=0$). In (\ref{hairy}), $N$ is a sort of {\it hair}. This hair is, however, of gravitational origin. One can easily verify that the value of $N$ explicitly appears in the scalar curvature and it makes the black hole with $N\neq 0$ to have a curvature singularity at $r=0$. It is also worth mentioning that a rotating generalization of (\ref{hairy}) also exists \cite{Julio2}. More interestingly, the solution above may also describe a static asymptotically de Sitter (dS) black hole, which follows from considering in (\ref{hairy}) the change $\ell^2\to -\ell^2$. In that case, $r_-$ gives the location of the black hole horizon, while $r_+$ is the radius of the cosmological horizon, which turns out to be in thermal equilibrium with the black hole. This also represents a solution to field equations (\ref{Prima}) in the limit $m^2\to 0$, $\mu \to 0$.

\section{Concluding remarks}

In this paper, we have continued the study of the vacua of the exotic massive 3D gravity theory \cite{Ozkan:2018cxj} initiated in reference \cite{Chernicoff:2018hpb}. We mainly focused our attention on asymptotically AdS$_3$, Warped-AdS$_3$, and AdS$_2\times S^2$ geometries. This includes black holes and non-linear gravitational waves. In the case of asymptotically AdS$_3$ geometries, we obtained the values of the central charge of the dual CFT$_2$ for the Cardy formula (\ref{Cardy1}) to reproduce the entropy of BTZ black holes of EMG. The theory, however, contains a much more rich bestiary of black holes with different asymptotics \cite{Chernicoff:2018hpb}; this includes Warped-WAdS$_3$ black holes, which we also discussed, together with Lifshitz black holes. In those cases, a similar type of holography inspired computation is in principle possible. Of particular interest is the case of Lifshitz black holes, for which EMG enables an arbitrary value of dynamical exponent $z$. It would be interesting to verify that a computation similar to the one in \cite{Hernan} can be done in EMG; see also \cite{Ayon-Beato:2015jga}. This would first demand to compute the conserved charges in non-constant curvature, asymptotically Lifshitz spaces for such a third-way gravity model. 

We also discussed deformations of AdS$_3$ and WAdS$_3$ black holes. In the case of AdS$_3$ space, we constructed a family of logarithmically decaying gravitons on the extremal BTZ black holes. These solutions are analogous to the Log-gravity solutions that TMG exhibits at the chiral point. In the case of WAdS$_3$ spaces, we gave explicit examples of evanescent gravitational waves propagating on massless black holes. We also discussed the conformal gravity limit of the theory and its locally AdS$_2 \times S^1$ solutions, recovering some results of \cite{AdS2} as a limit of the WAdS$_3$/CFT$_2$ symmetry analysis. Finally, we showed that, in a particular limit of the parameter space, EMG admits asymptotically hairy (A)dS$_3$ black holes with non-vanishing conserved charges and non-trivial thermodynamical properties. The fact that all these holographically well-motivated geometries appear in different corners of the space of solutions of EMG manifestly shows that this theory is a very nice toy model to explore holography beyond AdS. The theory, however, has peculiar features that make its holographic application knotty: The lack of an action in its metric formulation and the subtle problem of how to couple matter in a consistent way are two features that require further analysis. In relation to that, it would be interesting to investigate further the definition of EMG in the Chern-Simons like formulation, for which an action in terms of the dreibein and the spin connection is available. 

\section*{Acknowledgments}

The authors thank M. Chernicoff, N. Grandi, R. Mann, and S. Sajadi for discussions. This work was partially supported by FONDECYT grant number 1181047.

\end{document}